\documentclass[sigconf]{acmart}
\acmConference[MSR 2026]{MSR '26: Proceedings of the 23rd International Conference on Mining Software Repositories}{April 2026}{Rio de Janeiro, Brazil}

\usepackage{listings}
\usepackage{mdframed}

\definecolor{findings}{HTML}{28784D}

\newcommand{\RedL}{\makebox[0pt][l]{\color{red!6}\rule[-3pt]{\linewidth}{10pt}}}
\newcommand{\GreenL}{\makebox[0pt][l]{\color{green!6}\rule[-3pt]{\linewidth}{10pt}}}
\definecolor{green}{RGB}{0,127,0}

\lstdefinelanguage{diff}{
  basicstyle=\scriptsize\ttfamily \color{black},
  columns=fullflexible,
  breaklines=true,
  breakatwhitespace=false,
  showspaces=false,               
  showstringspaces=false,  
  frame=single, 
  showtabs=false,
  numbersep=5pt,
  showstringspaces=false,        
  stepnumber=1,                   
  tabsize=5,                     
  title=\lstname,  
  numbers=left,                 
  numbersep=5pt,   
  backgroundcolor=\color{white},
  morecomment=[f][\lstbg{red!5}]-,
  morecomment=[f][\lstbg{green!5}]+,
  morecomment=[f][\textit]{@@},
}

\copyrightyear{2026}
\acmYear{2026}
\setcopyright{cc}
\setcctype{by}
\acmConference[MSR '26]{23rd International Conference on Mining Software Repositories}{April 13--14, 2026}{Rio de Janeiro, Brazil}
\acmBooktitle{23rd International Conference on Mining Software Repositories (MSR '26), April 13--14, 2026, Rio de Janeiro, Brazil}
\acmPrice{}
\acmDOI{10.1145/3793302.3793563}
\acmISBN{979-8-4007-2474-9/2026/04}
\begin{document}

\title{AI builds, We Analyze: An Empirical Study of AI-Generated Build Code Quality}

\author{Anwar Ghammam}
\email{aghammam@umich.edu}
\affiliation{%
  \institution{University of Michigan-Deaborn}
  \city{Dearborn}
  \state{Michigan}
  \country{USA}
}

\author{Mohamed Almukhtar}
\email{almukhtr@umich.edu}
\affiliation{%
  \institution{University of Michigan-Flint}
  \city{Flint}
  \state{Michigan}
  \country{USA}
}

\renewcommand{\shortauthors}{Ghammam et al.}

\begin{abstract}The rapid adoption of AI coding agents for software development has raised important questions about the quality and maintainability of the code they produce. While prior studies have examined AI-generated source code, the impact of AI coding agents on build systems—a critical yet understudied component of the software lifecycle—remains largely unexplored. This data mining challenge focuses on AIDev, the first large-scale, openly available dataset capturing agent-authored pull requests (Agentic-PRs) from real-world GitHub repositories. Our paper leverages this dataset to investigate (RQ1) whether AI coding agents generate build code with quality issues (e.g., code smells), (RQ2) to what extent AI agents can eliminate code smells from build code, and (RQ3) to what extent Agentic-PRs are accepted by developers. We identified 364 maintainability- and security-related build smells across varying severity levels, indicating that AI-generated build code can introduce quality issues—such as lack of error handling, and hardcoded paths or URLs—while also, in some cases, removing existing smells through refactorings (e.g., Pull Up Module and Externalize Properties). Notably, more than 61\% of Agentic-PRs are approved and merged with minimal human intervention.
This dual impact underscores the need for future research on AI-aware build code quality assessment to systematically evaluate, guide, and govern AI-generated build systems code.


\end{abstract}





\maketitle

\section{Introduction}

The recent advances in modern software systems have enabled their rapid adoption of Large language Models and AI coding agents, and they are widely integrated into software development workflows \cite{grewal2024analyzing,nam2024using,coignion2024performance,fakhoury2024llm,yeticstiren2023evaluating,becker2023programming}. A growing body of research has examined the quality, correctness, and maintainability of AI-generated source code across multiple domains \cite{vaithilingam2022expectation,pearce2025asleep,asare2023github,siddiq2022securityeval,liu2024refining, siddiq2024quality, grewal2024analyzing}. However, while AI agents' assistance has been extensively studied for source code, comparatively little attention has been given to their emerging role in build systems. Modern software development relies heavily on build tools such as Maven \cite{Maven}, Gradle \cite{Gradle}, and Cmake \cite{swidzinski2022modern} to manage dependencies, compilation, packaging, and deployment, yet the quality of AI-generated build code remains largely unexplored. In particular, we lack empirical evidence on whether AI agents introduce or remove build-specific code quality issues and how developers respond to automated build code generation in practice.
We narrow this gap by conducting the first empirical study of AI-generated build code assessment across common build systems such as Maven, Gradle, CMake, and Make. Specifically, we focus on the following research questions:

\textbf{RQ1: Do AI-coding agents generate build code with smells?}
A recent study \cite{siddiq2022empirical} showed that datasets used to train open-source AI coding agents contain code and security smells that leak to the output generated by these models. In this RQ, we investigate whether these agents also generate build code with quality issues. Across 387 PRs and 945 build files, we found that AI agents mainly introduce maintainability-related code smells (e.g, Deprecated Dependencies, and Lack of Error Handling) and security-related code smells (e.g., Hardcoded Credentials).

\textbf{RQ2: To what extent can AI coding agents remove code smells from Build code?}
We examined whether AI changes reduce existing code issues in build scripts. We identified 31 build files where smells were eliminated, and a manual inspection showed that many improvements stemmed from refactoring changes, such as Pull Up Module, and Remove unused dependencies, which align with build-specific maintenance patterns reported in prior research on build refactoring \cite{ghammam2025build}. 

\textbf{RQ3: To what extent are Agentic-PRs accepted by developers?}
We examine whether developers accept AI-generated build-related code and analyze any modifications they make prior to merging. Our findings show that most AI-generated build PRs are merged with little to no manual adjustment, suggesting an early trust and adoption among developers.

The main contributions of our paper are as follows:
\begin{itemize}
    \item The first empirical study to assess the quality of AI-generated build code.
    \item Identification of refactoring actions performed by AI agents that help remove code smells in build scripts.
    \item An open-source dataset of labeled pull requests, including associated code smells and introduced refactorings, along with all manual annotations and fully reproducible code, is available in our replication package \cite{reppackage}.
\end{itemize}

\vspace{2.5em}
\section{Background}

\subsection{Build Systems}

Build automation is a critical part of modern software development \cite{mcintosh2012evolution}. Build tools provide flexible ways to model, manage, and maintain complex software projects. They rely on configuration files to define dependencies, specify build tasks, and orchestrate build process workflows \cite{Muschko_2014,Gradle,Ant,Maven,mcintosh2012evolution,clemencic2012cmake}. Within this work, we specifically focus on Maven \cite{Maven}, Gradle \cite{Gradle}, Cmake \cite{clemencic2012cmake} and Makefile \cite{makefile}, widely used build systems that provide a good representation of the different approaches to build automation.
\vspace{-0.8em}
\subsection{Code Smells \& Refactoring}
A code smell is an indicator of an improper choice of system design and implementation strategy \cite{CodeSmell}. These flaws can stifle software development or increase the chance of future errors \cite{reis2020code,tamanna2025your}.  Refactoring is defined as the process of restructuring existing code without changing its external behavior to improve readability, maintainability, and extensibility \cite{10.1145/2507288.2507326}. Refactoring can help reduce smells by improving code design and quality \cite{suryanarayana2014refactoring,ghammam2025build,ivers2024mind,almukhtar2025code}. In the context of build code, refactoring can involve simplifying build scripts, removing duplication, and improving maintainability \cite{duncan2009thoughtworks, ghammam2025build}.
\vspace{-0.8em}
\subsection{Sniffer: A Static Analyser for build code}
\label{sub:sniffer}
Sniffer \cite{tamanna2025your} is a static analysis tool specifically designed to detect code smells in build systems such as Maven, Gradle, CMake, and Make. Sniffer provides automated detection of both maintainability-related and security-relevant code smells, including \textit{outdated or inconsistent dependency declarations, Hardcoded Dependencies, insecure URLs, hardcoded credentials, Suspicious Comments, and overly complex build logic}. Sniffer analyzes each build file independently and reports build-specific smells along with their severity (low, medium, or high), a textual description, and the location of the issue. A comprehensive catalog of the code smells categories identified by Sniffer is provided in \cite{tamanna2025your,SnifferCodeSmell}, along with high-level descriptions of their characteristics.

\section{Related Works}
\subsection{Build Maintenance}
As software systems evolve, recent research has highlighted the importance of build maintenance as they get more complex. Hardt et al.~\cite{hardt2013ant} have introduced Formiga, a tool that facilitates build maintenance for Ant. Simpson et al.~\cite{duncan2009thoughtworks} have discussed the need for Ant code quality enhancement and lists 23 types of refactoring changes. Xiao et al.~\cite{xiao2021characterizing} identified a taxonomy of technical debts in Maven systems, mainly related to library limitations and dependency management issues. Similarly, Tammana et al. \cite{tamanna2025your} present the first study of code smells in build scripts across Maven, Gradle, CMake, and Make. The authors identify build-specific smell categories and
show that such smells are widespread and vary across build systems.
Ghammam et al. \cite{ghammam2025build} complement these findings by introducing a taxonomy of build refactorings in build systems and their mitigation of technical debt (e.g., Maintainability, build time)

\subsection{AI-based Code Generation}
Prior work investigated the usability of AI code generation \cite{vaithilingam2022expectation} and whether it generates vulnerable code \cite{pearce2025asleep}. Siddiq et al. \cite{asare2023github} focused on quality issues in the code generation datasets and their leakage from the closed and open source models. In another
study, authors created a framework to systematically evaluate generated AI code from a security perspective \cite{siddiq2022securityeval}. Liu et al. \cite{liu2024refining,siddiq2024quality,grewal2024analyzing} focused on the quality issues of ChatGPT-generated code. All of these related works are limited to source code, and none evaluate the build code quality generated by AI agents, which is the purpose of our study.

\section{Methodology}
\subsection{Dataset}
We base our analysis on the AIDev dataset \cite{li2025rise}, a large-scale dataset of around 933k agentic PRs spanning real-world GitHub repositories and generated by five AI agents: OpenAI Codex, Devin, GitHub Copilot, Cursor, and Claude Code. For the purposes of this study, we focus specifically on PRs that involve code changes to build systems—namely Gradle, Maven, CMake, and Make—as these represent the most widely used build technologies in modern software development \cite{maudoux2018correct}, and are supported by the static analyzer Sniffer \cite{SnifferCodeSmell}. In the following, we describe the steps taken to address our three research questions.
\vspace{-1em}
\subsection{Data Preprocessing}
To ensure that our dataset captures code changes related to build code, we filtered the AIDev dataset based on the modified files in each agentic PR. We kept only PRs that changed at least one build file across the studied build systems, such as Maven (XML-based pom.xml), Gradle (.gradle and .gradle.kts), CMake (CMakeLists.txt and .cmake files), and Make (Makefile and .mk). This initial filtering produced 632 candidate PRs. Through a subsequent manual inspection, we noticed several sources of noise: Some changed build files were hidden (e.g., referenced in .gitignore files); certain PRs were false positives where several extensions did not correspond to build scripts. Therefore, we applied a second round of manual filtering done by one of the authors to remove misclassified PRs, retaining only true build-related PRs. We obtained a final dataset of 387 PRs containing authentic build-system changes authored by AI agents. These consisted of 173, 104, 69, 38, and 2 PRs generated by Codex, Copilot, Devin, Cursor, and Claude Code, respectively.
\vspace{-1em}
\subsection{Build Code Quality Assessment With Sniffer}
To identify code smells, Sniffer \cite{SnifferCodeSmell} analyzes each build file independently. First, we started by extracting all changed or newly created build files from the studied PRs, yielding a total of 945 files. Next, we retrieved each file’s content both before and after the AI-generated change, obtaining paired snapshots that capture the exact AI-changes applied to the build code. Each pair was then analyzed with Sniffer to identify the presence and severity of code smells in both versions. Finally, we compared them to determine whether a given AI-change introduced new smells, eliminated existing ones, or produced no observable change. This comparison enabled us to quantify the effect of the AI-generated changes on build quality and assess whether they tend to improve, degrade, or preserve build-systems quality overall.

\subsection{Qualitative Analysis via Manual Labeling}
To answer RQ2, we focused on build-related changes in which code smells were eliminated. Two authors individually examined the build code changes, their associated commit history, and—when available—developers' comments and discussions. The goal was to determine how the eliminated smells were mitigated and whether quality-enhancing changes were introduced as part of the PRs. Once we completed the initial labeling phase, we measured the inter-rater agreement between the labelers using Fleiss’ Kappa \cite{campbell2013coding}, to evaluate both (i) the identification of quality-enhancing changes and (ii) their categorization into specific types. The resulting score of 0.76 indicates substantial agreement between the annotators \cite{campbell2013coding}. Finally, the two labelers conducted a follow-up consensus meeting to discuss and resolve remaining dissagreements.

To address RQ3, we investigated every PR—including those that introduced code smells, those that eliminated them, and those that had no observable impact on build quality—to determine whether the AI-generated changes were ultimately accepted and merged into the host repository. We examined the complete commit history of each PR, including any developer-authored revisions performed prior to merging. We also analyzed developers' discussions, review comments, and requested changes, if any, to understand how developers evaluate or refine AI-generated build code throughout the review process and before integration.



\vspace{-1em}

\section{Results}

\subsection{RQ1: AI-generated Build code Quality}
\label{sec:RQ1}
Our analysis of AI-generated build changes using Sniffer shows that the majority of the build files (824 out of 945) did not introduce any smells either before or after the Agentic changes. To better understand this "no smell" set, we manually inspected a representative sample (100 files). The manual analysis suggests that, in most of the cases, the agents applied simple changes (e.g., version updates or minor structural edits), which minimizes the opportunity to remove existing issues or the risk of introducing new smells. Meanwhile, in 66 build files, new smells were introduced following the AI-generated code, whereas 31 files were improved, with previously existing smells being removed. Finally, a total of 24 files are classified as neutral, where smells were present initially before the agentic changes and persisted after, indicating neither improvement nor degradation in build code quality.


\textbf{\textit{Introduced Build-specific Code Smells:}}
Across the 66 files, in which AI-generated changes introduced code quality issues, we found a total of 364 code smells with different levels of severity, as shown in Table \ref{tab:smell-distribution}. The distribution of detected code smells reveals that maintainability issues dominate AI-generated build changes. The most common issue, \textit{Wildcard Usage} (97 occurrences with medium severity), stems from using dependency version patterns like “+” or “*”, which can cause unpredictable build outcomes. \textit{Lack of Error Handling} occurs 63 times with medium severity, pointing to challenges in how AI agents ensure build reliability. Also, \textit{Deprecated Dependencies} and \textit{Outdated Dependencies} (both with medium to high severity) appear 56 and 27 times, respectively, indicating widespread reliance on outdated or unmaintained libraries. Other maintainability-related smells—such as \textit{Suspicious Comments} (23), \textit{Duplicate Code/Dependency} (15), and \textit{Complexity} (27)—appear but less often and generally reflect smaller organizational or readability concerns rather than serious risks.
On the security side, the code smell \textit{Hardcoded Paths/URLs} was frequently observed (25 occurrences) and ranged from minor to severe depending on the context: 13 low, 9 medium, 1 low-medium, and 2 high-severity. While \textit{Insecure URLs} was rare (only 1 high-severity occurrence), which is reassuring, even a few occurrences warrant attention. 



\begin{table}[ht!]
\vspace{-1em}
\centering
\caption{Distribution of Detected build-related Code Smells}
\label{tab:smell-distribution}
\footnotesize
\begin{tabular}{r p{2.2cm} p{4cm}}
\hline
\textbf{Count} & \textbf{Severity} & \textbf{Code Smell} \\
\hline

97  & Medium       & Wildcard Usage \\
63  & Medium       & Lack Error Handling \\
56 & Medium--High & Deprecated Dependencies \\
27  & Low          & Complexity \\
27  & Medium--High & Outdated Dependencies \\
23  & Low          & Suspicious Comments \\
19  & Medium       & Missing Version in Raw \\
15  & Low          & Duplicate (Code/Dependency) \\
13  & Low          & Hardcoded Paths/URLs \\
9   & Medium       & Hardcoded Paths/URLs \\
5   & High         & Hardcoded Credentials \\
4   & Low          & Deprecated Dependencies \\
2   & High         & Hardcoded Paths/URLs \\
2   & High         & Insecure URLs \\
1   & Low--Medium  & Hardcoded Paths/URLs \\
1   & Medium       & Inconsistent Dependency Management \\
\hline
\end{tabular}
\vspace{-1em}
\end{table}

\begin{figure}[ht!]
\vspace{-1em}
    \centering
    \includegraphics[width=1\linewidth]{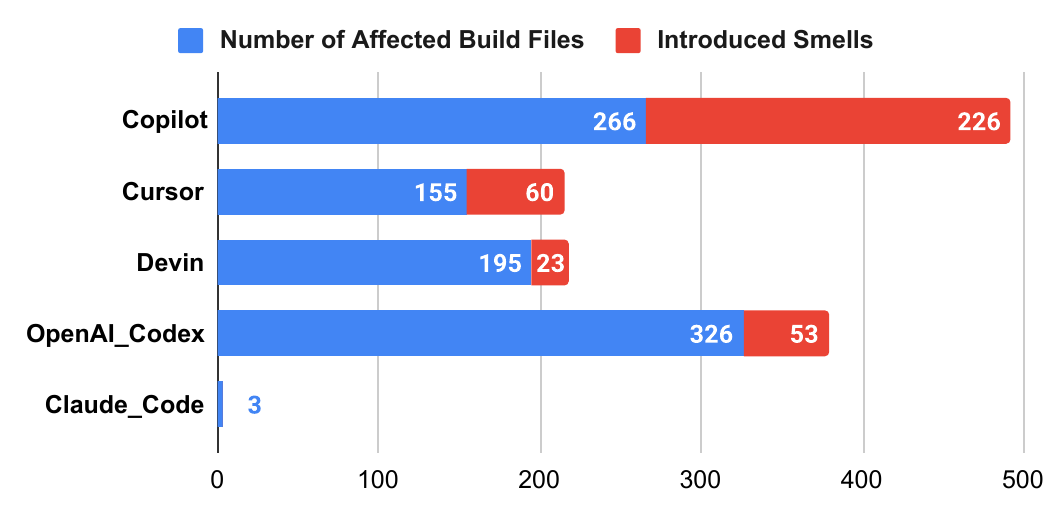}
    \caption{Introduced Code Smells per AI Agent}
    \label{fig:smells-per-agent}
\end{figure}

Figure \ref{fig:smells-per-agent} shows clear differences in how often AI coding agents introduce build-related code smells. Copilot introduces the most smells overall (226 smells across 266 build files), giving it the highest smell-introduction rate. Cursor follows with 60 smells across 155 files. OpenAI Codex, despite modifying the largest number of files (326), introduces far fewer smells (53). Devin also has a relatively low rate, with 23 smells across 195 files. Finally, Claude Code introduces none, though this result is based on only three build files. These differences highlight that AI assistants exhibit distinct behavioral patterns when modifying build code, and some agents may require stronger guardrails or integrated quality checks to prevent degradation of build quality.

\vspace{-2em}
\subsection{RQ2: Drivers of Code-Smells Elimination}
\label{sec:RQ2}
As discussed in Section \ref{sec:RQ1}, 31 build files showed quality improvement following the AI-generated changes. Across them, a total of 54 code smells were removed. The resolved smells included 7 wildcard usages, 6 outdated dependencies, 5 hardcoded paths/URLs, 5 duplicate code or dependency blocks, 3 hardcoded credentials, 4 suspicious comments, 2 deprecated dependencies, 2 insecure URLs, and 1 inconsistent dependency. We applied a manual analysis of the affected files to understand how these smells were resolved; Results show that about 65\% of the eliminated smells resulted from intentional, quality-oriented changes. Across them, the AI agents performed \textbf{\textit{Reformat Code}} (3 instances), \textbf{\textit{Remove Unused/Redundant code}} (7 instances), \textbf{\textit{Extract and Pull Up Variable}} to address reusability and inconsistent dependencies (1 instance), \textbf{\textit{Pull Up Module}} used to centralize shared elements, which eliminates hardcoded paths/URLs and deprecated components (1 instance), and \textbf{\textit{Externalize Properties}} (1 instance) to remove hardcoded URLs and credentials.
These quality-enhancing code changes align closely with established refactoring patterns reported by Ghammam et al.\cite{ghammam2025build}, which demonstrated that developers actively employ refactoring changes in build code to mitigate quality issues in build systems. 
Listing \ref{lst:extractmethod} lists an example of the \textit{Externalize Properties} refactoring applied by Copilot in a build.gradle file. In this PR, hard-coded urls are replaced with dynamic values, allowing deployment paths to be determined at build time. Credentials that were previously embedded directly in the script are also moved into project-level properties, ensuring that sensitive information is no longer stored within the build file. This refactoring removes security-critical and hard-coded data, increases security and portability, and enhances long-term maintainability as publishing targets evolve. Copilot’s commit message documents the intention behind the change (e.g., “Maintained existing URLs for transition compatibility; preserved existing credentials and env variables”).

A second example appears in the PR from \cite{PR1}, where Codex applies an \textit{Extract and Pull-Up Variable} refactoring. In this case, a deprecated hard-coded dependency version is replaced with a parameter defined in the parent dependencies.gradle file, thereby centralizing version management and eliminating an Inconsistent Dependency Management smell.



\subsection{RQ3: To what extent are Agentic-PRs accepted by developers? }
Overall, 61.4\% (238/387) of the analyzed PRs were merged by developers, and in most cases, the merge occurred immediately after automated checks passed, with no additional modifications from human reviewers. Reviewers often accept the PR changes with a simple LGTM (Looks Good To Me) before integration (e.g., PR \cite{PR4}, \cite{PR5}, \cite{PR6}). In several instances, they explicitly praised the agent-generated changes (e.g., “You are awesome! You can take my job now” in PR \cite{PR2}), while some reviewers simply issued new prompts with minor instructions without manually altering the PR content (PR \cite{PR3}). 
The high merge ratio, combined with direct positive feedback and minimal review edits, suggests that developers currently perceive AI-generated build changes as sufficiently efficient and trustworthy to integrate rapidly.

\section{Discussion}
\textbf{AI-Generated Build Code Quality.} Our findings in RQ1 show that, despite the high adoption rate of AI-generated code nowadays, we note it may occasionally introduce maintainability-related smells (e.g., Deprecated dependencies, Wildcard Usage, Lack of Error Handling, etc.) and security-related smells (e.g., Hardcoded Credentials and Insecure URL) in build systems. While RQ2 findings show that in many cases, AI-generated code effectively removes smells through recognized refactoring changes. This highlights the need for developer oversight when integrating AI-generated build updates and points to an important direction for future work: improving how models are trained or guided so that they produce consistent, high-quality refactorings rather than unintentionally introducing technical debt through redundant or inconsistent build configurations.
\vspace{-1em}
\begin{center}
\begin{lstlisting}[language=diff,caption={PR \textit{\textsl{3483}} in \textit{\textsl{b1c-syntax/bsl-language-server}} Project: An Example of Copliot-generated \textbf{Externalize Properties} Refactoring in Gradle.},label={lst:extractmethod},numbers=left,frame=lines, escapechar=\!]
!\RedL! url = if (isSnapshot)
!\RedL!     uri("https://s01.oss.sonatype.org/content/repositories/snapshots/")
!\RedL!     else uri("https://s01.oss.sonatype.org/service/local/staging/deploy/maven2/")
!\RedL!   credentials { username = sonatypeUsername password = sonatypePassword}
!\GreenL! url = uri("${layout.buildDirectory.get()}/staging-deploy")}
\end{lstlisting}
\end{center}

\textbf{Implication for Developers.} AI agents are generally well-received by developers and are frequently merged with limited manual intervention. The high merge ratio, combined with direct positive feedback and minimal review edits, suggests that developers currently perceive AI-generated build changes as sufficiently efficient and trustworthy to integrate rapidly. Consequently, while AI agents clearly support developer build workflows, the AI-generated build code should be appropriately vetted to consistently align with established best practices rather than propagating new smells or technical debt in build systems.

\section{Threats To Validity}
In our work, the dataset is used for a mining challenge, AIDev dataset \cite{li2025rise}, which can introduce external threats to validity. However, the dataset is vetted by the organizers. Our work focused on build systems code, namely Gradle, Maven, CMake, and Make, as these represent some of the most widely used build technologies in modern software development \cite{maudoux2018correct}.

Another validity threat to this work is that we use Sniffer, a static
analyzer to detect code quality issues (code smells). Sniffer can introduce false positives, but their precision is significantly high according to the literature \cite{SnifferCodeSmell}. Another threat to this work concerns the manual evaluation of the PRs’ discussions and chats. The author's professional software development experience and the peer review mitigated this internal validity threat.

\section{Ethical Considerations}
Our study uses data from public GitHub repositories included in the AIDev dataset released for the MSR Mining Challenge. We analyze and report results only at an aggregate level, and our dataset does not include private or personal data.

\section{Conclusion \& Future Work}
AI agents are nowadays widely used in modern software engineering and increasingly contribute to the generation of software build code. Our study shows that AI agents can meaningfully refactor build scripts and eliminate certain build smells, although they also occasionally introduce new ones. We further observed that most agent-generated build PRs are merged with minimal intervention, indicating early trust and adoption in practice. Going forward, our research will explore smell-aware and refactoring-guided AI agent behavior, extend analyses to additional build systems, and integrate automated quality checks to prevent the accumulation of code issues through automated build modifications.


\bibliographystyle{ACM-Reference-Format}
\bibliography{ref}


\end{document}